\documentclass[lettersize,journal]{IEEEtran}
\usepackage{amsmath,amsfonts}
\usepackage{algorithmic}
\usepackage{algorithm}
\usepackage{array}
\usepackage[caption=false,font=small,labelfont=sf,textfont=sf]{subfig}
\usepackage{textcomp}
\usepackage{stfloats}
\usepackage{url}
\usepackage{verbatim}
\usepackage{color, soul}
\usepackage{graphicx}
\usepackage{cite}
\usepackage{xcolor}
\usepackage{tabularray}
\usepackage{placeins}
\usepackage{booktabs}
\usepackage{multirow}
\usepackage{tabularx}
\usepackage{adjustbox}
\usepackage{caption}
\usepackage{balance}
\usepackage{booktabs}
\usepackage{tikz}
\newcommand*\circled[1]{\tikz[baseline=(char.base)]{
            \node[shape=circle,draw,inner sep=1.3pt] (char) {#1};}}
\usepackage[left=1.62cm,right=1.62cm,top=0.7in]{geometry}
\begin{document}

\title{Volt/VAR Optimization in the Presence of Attacks: A Real-Time Co-Simulation Study}

\author{
\IEEEauthorblockN{\textbf{Mohd Asim Aftab}\IEEEauthorrefmark{1}, \textbf{Astha Chawla}\IEEEauthorrefmark{2},  \textbf{Pedro P. Vergara}\IEEEauthorrefmark{3}, 
\textbf{Shehab Ahmed}\IEEEauthorrefmark{1}, \textbf{Charalambos Konstantinou}\IEEEauthorrefmark{1}}

\IEEEauthorblockA{
\IEEEauthorrefmark{1}CEMSE Division, King Abdullah University of Science and Technology (KAUST)\\
\IEEEauthorrefmark{2}Siemens, India\\
\IEEEauthorrefmark{3}Intelligent Electrical Power Grids, Delft University of Technology (TU Delft)
}

\IEEEauthorblockA{
E-mail: \{mohammad.aftab, shehab.ahmed,  charalambos.konstantinou\}@kaust.edu.sa}, \\astha.chawla@siemens.com, p.p.vergarabarrios@tudelft.nl

}

\IEEEaftertitletext{\vspace{-2.2\baselineskip}}

\maketitle

\begin{abstract}
Traditionally, Volt/VAR optimization (VVO) is performed in distribution networks through legacy devices such as on-load tap changers (OLTCs), voltage regulators (VRs), and capacitor banks. With the amendment in IEEE 1547 standard, distributed energy resources (DERs) can now provide reactive power support to the grid. For this, renewable energy-based DERs, such as PV, are interfaced with the distribution networks through smart inverters (SIs). Due to the intermittent nature of such resources, VVO transforms into a dynamic problem that requires extensive communication between the VVO controller and devices performing the VVO scheme. This communication, however, can be potentially tampered with by an adversary rendering the VVO ineffective. In this regard, it is important to assess the impact of cyberattacks on the VVO scheme. This paper develops a real-time co-simulation setup to assess the effect of cyberattacks on VVO. The setup consists of a real-time power system simulator, a communication network emulator, and a master controller in a system-in-the-loop (SITL) setup. The DNP3 communication protocol is adopted for the underlying communication infrastructure. The results show that corrupted communication messages can lead to violation of voltage limits, increased number of setpoint updates of VRs, and economic loss. 
\end{abstract}
\vspace{-5mm}

\vspace{-3mm}
\section{Introduction}
Electric power distribution systems currently adopt Volt/VAR optimization (VVO) as a method to increase energy and operational efficiency as well as reduce power losses through voltage control. Specifically, the distribution network experiences a voltage drop as one moves from the substation to the remote feeder end. This problem is typically resolved by employing VVO algorithms in the distribution network to realize voltage regulation as well as loss reduction. VVO can provide setpoints to on-load tap changers (OLTCs), voltage regulators (VRs), and shunted capacitor banks to achieve desired voltage profiles. These components are scheduled to maintain voltage within the specified limits through day-ahead scheduling based on load forecasting. The main objective of VVO is to achieve conservation voltage reduction (CVR). CVR, which has received significant attention in recent literature \cite{VVOSurvey_9693496}, is a technique at the distribution system operator (DSO) level to deliberately reduce the feeder voltage and maximize energy savings. The voltage is maintained within the lower bounds of ANSI standard C84.1-2011 ($0.95$-$1.05$ per unit (p.u.)) \cite{Wang_CVR_review_6670787}. Annual energy savings by implementing CVR are in the range of  0.5\% to 4\% \cite{CVR}.

In recent years, increased attention towards environment-friendly energy generation led to the penetration of renewable energy resources (RES) in distribution networks. RES-based distributed energy resources (DERs), such as rooftop solar PV, are interfaced with the distribution networks through inverters which are responsible for the DC/AC conversion with a unity power factor. However, the revised IEEE 1547-2018 standard allows DERs to participate in grid support functions such as Volt/VAR and Volt/Watt control\cite{IEEE1547_9849493}. Thus, in modern distribution networks, DERs are paired with smart inverters (SIs) to support relevant functionalities per the recommendation of IEEE 1574-2018 standard.  


Due to the intermittent nature of renewable-based DERs, the VVO problem transforms into a dynamic problem in modern distribution networks. Its solution requires coordinated operation among OLTCs, VRs, shunted capacitor banks, and SIs, which are collectively termed VVO actors. The VVO actors extensively communicate with the controller to realize the VVO scheme.  
Measurements from smart meters and micro-PMUs (phasor measurement units) are acquired and transmitted to the VVO controller to compute optimized setpoints for VVO actors to maintain system voltage. 
To support this operation, extensive communication is required to achieve VVO in modern distribution networks. As a result, such schemes can be susceptible to cybersecurity effects due to the reliance on information and communication technologies \cite{9351954}. The potential vulnerability to cyberattacks could lead to power quality issues in distribution networks as well as under/overvoltage conditions and even voltage instability events.

The relevant literature indicates that attacks on the communication infrastructure of energy systems can lead to  disastrous consequences \cite{ https://doi.org/10.1049/iet-cps.2017.0033, SurveyofCyberattacks8598991}. Considering malicious attacks on DERs and their supporting ecosystem \cite{zografopoulos2022distributed}, several strategies for mitigating the effects on VVO are reported in the literature. In \cite{BDD_VVO7959131}, the detection of bad data in critical measurements of PV inverters which can lead to erroneous VVO is reported. For the mitigation of cyberattacks, solutions based on local measurements and historical data are proposed. A tri-level defender-attacker-defender optimization model is proposed in \cite{CHOEUM_trilevelDAD_2021117710} to mitigate cyberattacks against VVO. In \cite{Anto_secureVVC8954624}, the authors present a secure autonomously coordinated VVO scheme in the presence of high DER penetration to avoid grid oscillations and unintended switching operations of voltage regulators. 
In \cite{9730037_adaptivesecureVVO}, an adaptive scheme is proposed to mitigate cyberattacks on a number of SIs in the distribution network. The presented technique  adjusts the setpoints of non-attacked SIs to maintain system-wide voltage stability. 
 
It is evident that existing literature is primarily focused on mitigating and defending against cyberattacks in distribution networks. 
However, the research is limited on analyzing such attacks on VVO schemes in real-time co-simulation testbeds and how vulnerabilities in the communication medium can be exploited to launch these attacks. Hence, to fill this gap, this paper develops a real-time co-simulation setup for an advanced distribution management system (ADMS) to assess cyberattack impacts on the VVO scheme. The system-in-the-loop (SITL) setup comprises of real-time power system simulator
, a network emulator
, and a master controller to emulate a DNP3 client and run the VVO algorithm. 
The setup allows to test the actual control code, control hardware, and communication infrastructure realistically as they would be utilized in the field, representing an accurate representation of the actual system, and providing insights on the severity assessment of cyberattacks in ADMS. 

The paper is organized as follows. Section \ref{s:problemformulation} presents the formulation of the VVO problem to achieve CVR. Section \ref{s:setup} explains the development of the real-time co-simulation setup for ADMS and  the threat model of cyberattacks on VVO scheme. Finally, Section \ref{s:results} provides the results of the impact assessment, while Section \ref{s:conclusions} concludes the paper.
 
\vspace{-2mm}
\section{VVO Problem Formulation for CVR}\label{s:problemformulation}
The VVO is a multi-objective nonlinear optimization problem aimed at maintaining the distribution network voltage profile. The objectives of VVO considered in this paper are the minimization of distribution network losses and energy savings using CVR. The VVO algorithm is solved using the alternating direction method of multipliers-based scheme (ADMM) \cite{VVO_solver}. The VVO algorithm computes the updated setpoints for VVO actors. The VVO formulation, in this paper, considers VRs, capacitor banks, and photovoltaic smart inverters (PVSIs). The objective function is formulated as the minimization of Eq. \eqref{1}: 
\begin{equation}
    F=Min\left[C_{loss,t_k}+C_{VR,t_k}+C_{CB,t_k}+C_{PVSI,t_k}\right]\label{1}
\end{equation}where $F$ is the objective function, $C_{loss,t_k}$ is the grid loss cost, and $C_{VR,t_k}$, $C_{CB,t_k}$, $C_{PVSI,t_k}$ are operating costs of VRs, capacitor banks, and SIs,  respectively.

The cost of power loss in the distribution network is computed as the product of the cost of energy and power loss in line resistance as shown below:
\begin{equation}
C_{loss,t_k}=C_{E}*{i^2_{i,j}r_{i,j}}\label{2}
\end{equation}where $C_{E}$ is the cost of energy, and ${i^2_{i,j}r_{i,j}}$ is the power loss between ${i_{th}}$ and ${j_{th}}$ node. 
Considering $R$ as the number of VRs, and $B$ as the number of capacitor banks in the distribution network, the cost of VRs is computed as the product of the cost incurred for VRs operation per step, ${C_{r,t_k}}$, and the modulus of change in the number of steps of VRs, $\Delta X_{VR}$, as per Eq. \eqref{3}. Similarly, the cost of the capacitor banks is computed as per Eq. \eqref{4}: 
\begin{equation}
{C_{VR,t_k}=\sum^R_{i=1}{C_{r,t_k}*|}\Delta X_{VR}|}\label{3}
\end{equation}
\begin{equation}
C_{CB,t_k}=\sum^B_{i=1}{C_{b,t_k}*|\Delta X_{CB}|}\label{4}
\end{equation}

In our problem, DERs are represented by PV power plants  interfaced with the distribution network through SIs. Consumers with PVSI facilities participate in VVO. The SI is operated in Volt/VAR mode to provide reactive power support to the distribution network. The cost of PVSI operation is computed as the incentives provided to PVSI owners per unit of reactive power supplied, ${Q^i_{PVSI}}$,  based upon the cost of grid power {at that specific time instant}, ${C_{grid,t_k}}$, as follows:
\begin{equation}
C_{PVSI,t_k}=\sum^P_{i=1}{C_{grid,t_k}*Q^i_{PVSI}}\label{5}
\end{equation}

The objective function of Eq. \eqref{1} is solved with certain constraints presented in Eqs. \eqref{6}-\eqref{10}. The voltage limits on the feeder must be maintained within limits as specified in \cite{Shailendra_Timebased_ANSI_8762192}. The operational constraint for every  ${i_{th}}$ bus is given as:  
\begin{equation}
\noindent 
0.95\ pu\le V_i\le 1.05\ pu \label{6}
\end{equation}
The VRs in substations are equipped with taps to participate in voltage control. However, they are limited by the tap's operational limits as per Eq. \eqref{7}: 
\begin{equation}
0.9\ pu\le {Tap}_{i,VR}\le 1.1\ pu \label{7}
\end{equation}

SIs can contribute to VVO by modulating their reactive power output as a function of voltage measured at the point of common coupling (PCC) \cite{Singhal_RealtimelocalVVO8365842}. In order to fully utilize their capability of reactive power support, the SIs apparent power rating is designed to be more than the active power rating. In this way, each SI can provide reactive power support even if its active power output is dispatched at maximum value \cite{IEEE2030NREL}. The PVSI must balance the relation between active and reactive power as per Eq. \eqref{8}. To follow the industry best practices, the PVSI is limited to delivering 60\% of its total rated capacity as per Eqs. \eqref{9}-\eqref{10}.
\begin{equation}
S^2_{PVSI}=P^2_{PVSI}+Q^2_{PVSI}\label{8}
\end{equation}
\begin{equation}
Q^i_{PVSI,t}={\beta }_{PVSI}*S^i_{PVSI,t}\label{9}
\end{equation}
\begin{equation}
0\le {\beta }_{PVSI}\le 0.6\label{10}
\end{equation}
\noindent 

\noindent where ${S_{PVSI}}$, ${P_{PVSI}}$, and ${Q_{PVSI}}$ are the apparent, active, and reactive power of SI, respectively.

In addition, the power flow equations for active and reactive power must be satisfied as presented in \cite{ShailendraEventdrivenPV_9350202}. 
For assessing the energy savings through CVR, the loads in the distribution network are modeled using the ZIP load model. This model expresses the correlation between voltage magnitude and power through a polynomial equation that incorporates constant impedance (Z), current (I), and power (P) components.
The VVO aims to reduce the grid's loss as well as operational expenses, taking into account network limitations at each quasi real-time phase. The VVO algorithm as in Eq. \eqref{1} is executed subject to Eqs. \eqref{6}-\eqref{10} and setpoints for VRs, CBs and PVSIs are obtained. 
These setpoints change with the load profile of the feeder and with variations in solar irradiance. The VVO is performed every 15 minutes, and new settings are communicated to the VVO actors to maintain the voltage profile of the feeder and maximize energy savings for the DSO. 

\begin{figure}[t]
\centering
\includegraphics[width=2.7in]{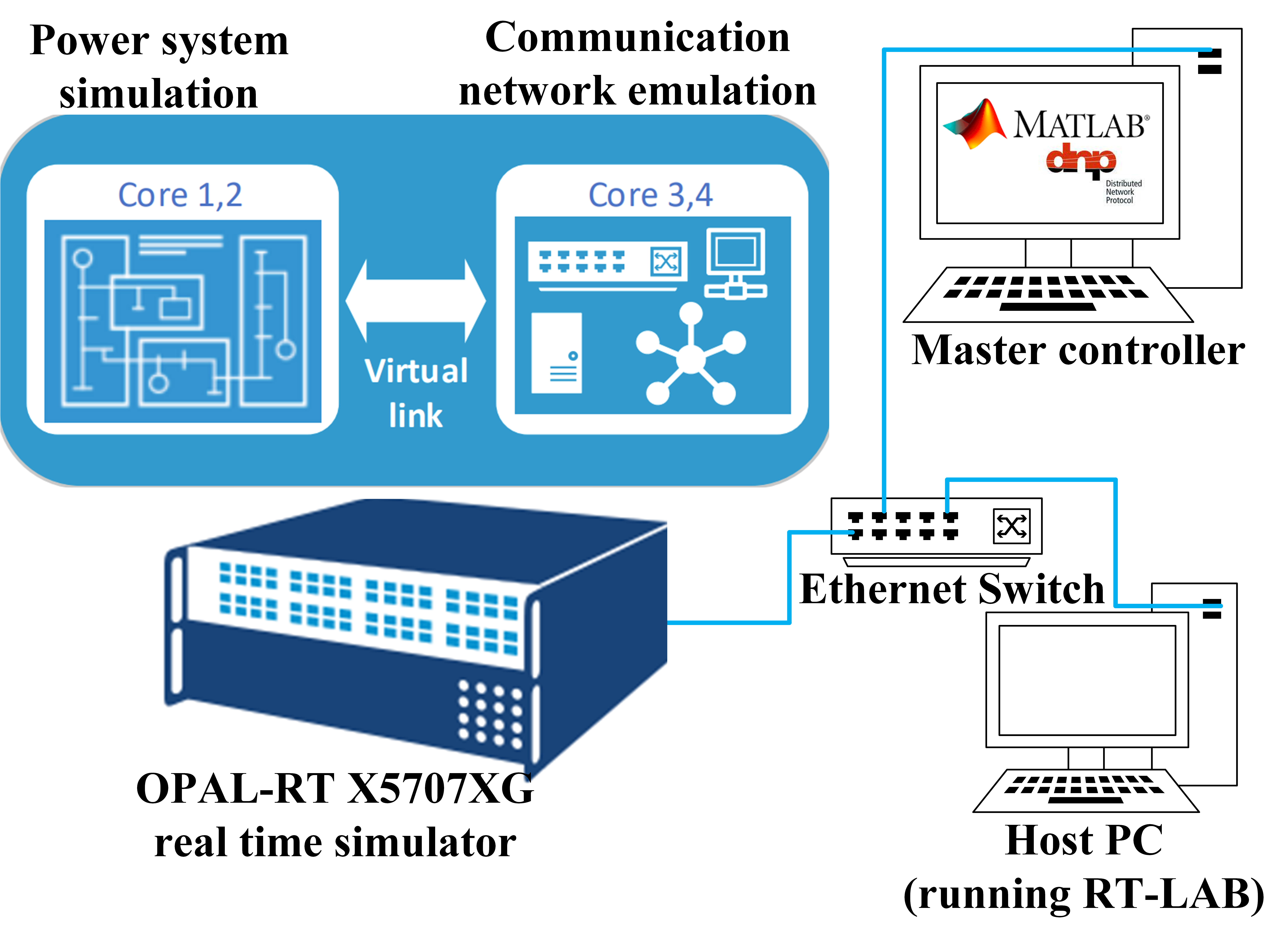}
\vspace{-1mm}
\caption{Overview of the developed advanced distribution management system-in-the-loop (SITL) co-simulation setup.}
\vspace{-4mm}
\label{fig_1}
\end{figure}

\vspace{-1mm}
\section{Real-time Co-Simulation Setup for ADMS}\label{s:setup}
The assessment of cyberattacks on ADMS in a real-time setup provides a realistic and controlled environment to evaluate different scenarios, identify vulnerabilities, and validate the effectiveness of security measures. It also ensures the reproducibility of results. For these reasons, a real time co-simulation setup employing both a real time simulator and a network emulator is utilized in this paper. 
\vspace{-1mm}

\subsection{Overview of the Co-Simulation Setup}
The real-time co-simulation setup for the ADMS consists of a real-time power system simulator (\textsc{OPAL-RT} OP5707XG), a communication network emulator (\textsc{ExataCPS}), and a master controller for monitoring and control. The setup is implemented over the DNP3 TCP/IP-based protocol, typically employed in distribution networks in North America. DNP3 is a client-server layer 2 communication protocol. It also defines generic data types employed for supervisory control and data applications. An overview of the ADMS SITL setup is shown in Fig. \ref{fig_1}. Different components of the setup are explained below.

\begin{figure}[t]
\centering
\includegraphics[width=2.8in]{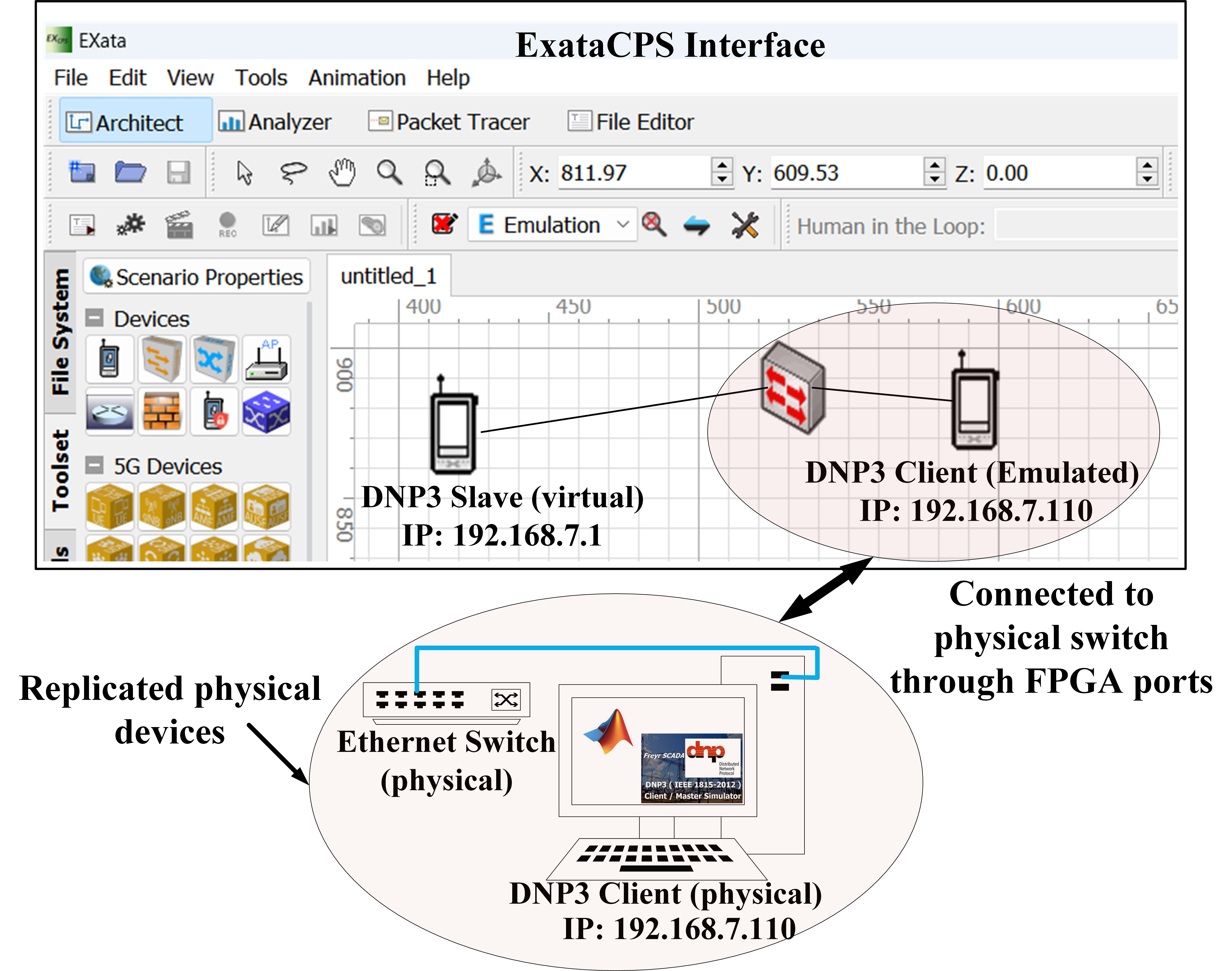}
\vspace{-1mm}
\caption{\textsc{ExataCPS} interface with the master controller.}
\vspace{-6mm}
\label{fig_2}
\end{figure}


The distribution network is modeled in a \emph{real-time power system simulator}, the \textsc{OPAL-RT} simulator (OP5707XG), which is an FPGA-based hardware. It can mimic the operation of power systems in real-time, ensuring that all iterations of the model are completed in a prescribed amount of time at each time-step. 


A \emph{communication network emulator} is used to replicate the behavior of the communication network with high accuracy. \textsc{ExataCPS} network emulator is employed for modeling the network of the distribution system. \textsc{ExataCPS} runs by reserving one of the FPGA cores on \textsc{OPAL-RT} and interacts with the power system simulation running on \textsc{RT-LAB} through a virtual link as shown in Fig. \ref{fig_1}. This is advantageous compared to other communication network emulators, which run in local machines and interact with real-time power system simulators through network interface cards (NICs). In such setups, the simulation time-step of the power system in a dedicated real-time simulator and the communication network emulation in the local machine are different. This issue is resolved by running \textsc{ExataCPS} on one of the cores of \textsc{OPAL-RT} with the exact same time-step. 


The \emph{master controller} is a PC running the DNP3 client software and VVO code in \textsc{MATLAB}. The DNP3 client simulator from \textsc{FreyrSCADA} is employed as the DNP3 client software. It sets up a DNP3 master connection through TCP/IP to the DNP3 slave running as a virtual device inside the \textsc{ExataCPS} emulation. The DNP3 analog input messages are saved in \texttt{.csv} format within \textsc{MATLAB}. An ADMM-based VVO code is used to run the optimization program and compute new setpoints for VRs and the Volt/VAR curve for the SI \cite{VVO_solver}. The DNP3 master sends analog output point commands to set the new values in the distribution network simulation through \textsc{ExataCPS} interface. The DNP3 client is an emulated intelligent electronic device (IED) that communicates to a virtual slave inside \textsc{ExataCPS} as shown in Fig. \ref{fig_2}.
\vspace{-1mm}

\vspace{-2mm}
\subsection{Threat Model and Attack Scenarios}
In order to assess the impact of cyberattacks on VVO in the developed setup, the threat model and attack scenarios are described below. The load measurements through smart meters located on nodes of the distribution network are periodically updated and transferred to the VVO controller. The VVO controller computes new setpoints for voltage control. With the inclusion of DERs through SIs, the DER owners also participate in the VVO scheme. These SIs are connected through public networks of DER owners who are not fully aware of cyber-threats and related potential vulnerabilities. Thus, adversaries can gain access control through DER units and launch cyberattacks. The primary goal of such attackers is to  
stealthily change the voltage level of the distribution feeder.  

In this work, we consider that the cyberattacks on VVO are realized via a data integrity modification \cite{zografopoulos2021security, intriago2023residual}. We consider malicious adversaries accessing and replacing real measurements \(y\) in the system with fake information \(\hat{y}\). This can be modeled as an optimization problem, where the adversary aims to minimize the detection of the attack while maximizing the damage caused, i.e., $\text{minimize} \left[f(\hat{y}) + \lambda \cdot g(\hat{y}, y) \right]$, where \(f(\hat{y})\) represents a cost function that quantifies the damage caused by the attack, and \(g(\hat{y}, y)\) represents a measure of the similarity between the tampered measurements \(\hat{y}\) and the original measurements \(y\). The parameter \(\lambda\) controls the trade-off between causing damage and avoiding detection. To perform the attack, the adversary manipulates the measurements by adding a perturbation vector \(d\) to the real measurements, i.e., $\hat{y} = y + d$. Specifically, we assume that an attacker is able to tamper DNP3 communication messages and modify exchange packets, i.e., \textit{modify packets (MODP)}, on the DNP3 client node. Data packet modification can be performed by adding on the offset, inverting data bits, multiplying each field with a value, replacing bits with random values, or even completely interrupt the communication (i.e., denial-of-service (DoS)). The DNP3 frame has a header and data section and the maximum frame length can be 292 bytes. For launching the MODP attack in \textsc{ExataCPS}, the starting byte, i.e., the byte at which the DNP3 payload begins, needs to be specified so that the attack produces ``correct'' results.  

\begin{figure}[t]
\centering
\includegraphics[width=3.3in]{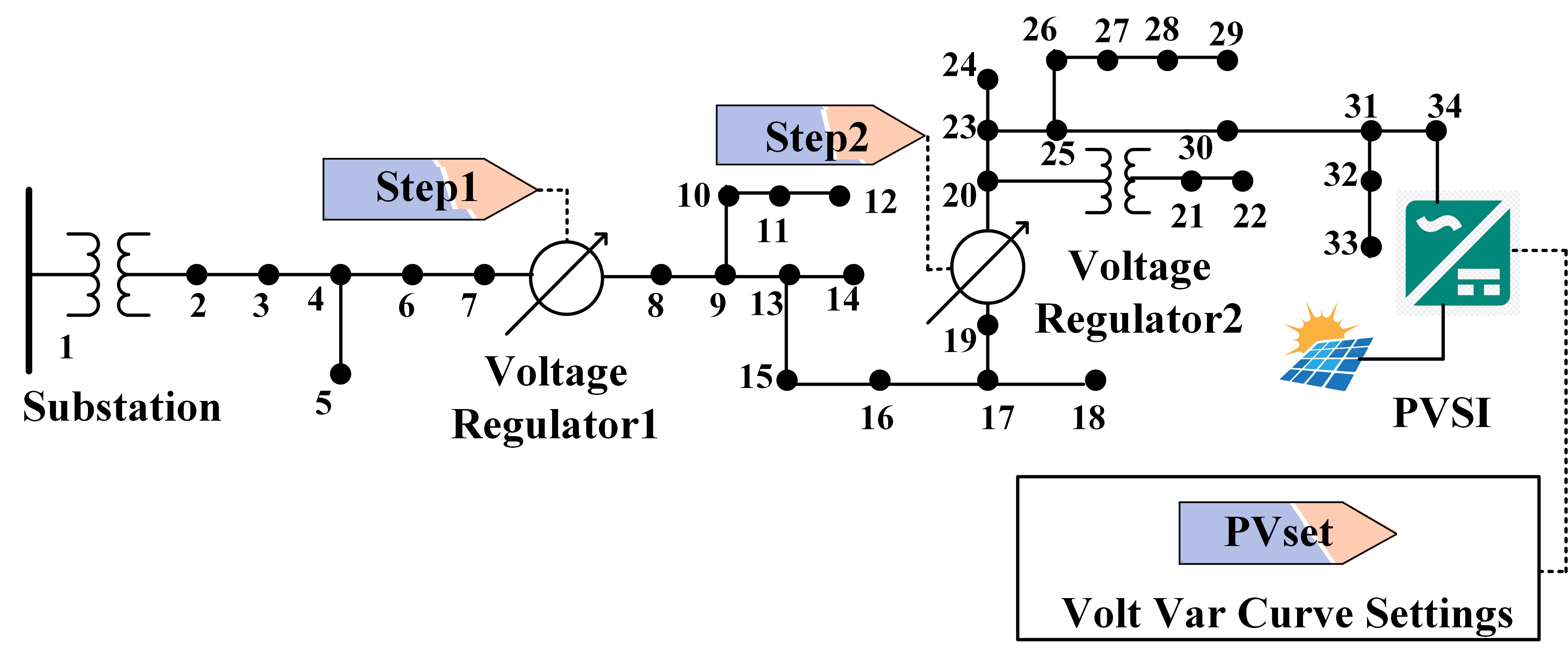}
\vspace{-1mm}
\caption{Schematic representation of modified IEEE 34-bus system.}
\vspace{-4mm}
\label{fig_3}
\end{figure}

In our experiments, the \emph{MODP attack is realized under three categories in terms of the considered VVO application} \cite{9146704}:

\noindent \circled{1} \textit{Tampering VVO setpoints}: 
The setpoints for VVO actors are communicated through DNP3 direct operate messages. These setpoints can be deliberately tampered via MODP by an attacker. As a result, conditions could arise related to \textit{(i)} under/overvoltage violations in the operating feeder, \textit{(ii)} exceeding operating voltage limits as per the ANSI standard C84.1-2011, and \textit{(iii)} economic loss to the DSO.
    
\noindent \circled{2} \textit{Tampering smart meter measurements}: In this scenario, tampering with active and reactive power measurements through a smart meter is considered. The primary goal of the attacker is to render the VVO scheme ineffective due to the corrupted  measurements. 
For this, a DoS attack can be launched by an attacker (e.g., cancel via a MODP perturbation vector \(d\) the real measurements $y$, so that $\hat{y}$ is nullified) which can result in the failure of communication infrastructure. 
    
\noindent \circled{3} \textit{Modifying SI Volt/VAR curve setpoints}: The attacker can modify the Volt/VAR curve of the SI either by changing its slope or by changing the voltage dead band \cite{9146704}. As a result, unwanted oscillations could arise in the distribution feeder. Under this attack scenario, and in order to maintain the voltage within limits, there can be an increased number of setpoint updates of VR's, leading to economic loss and even sometimes failure of the VVO controller to produce optimized setpoints, and hence rendering VVO ineffective. 

\begin{figure}[t]
\centering
\includegraphics[width=3.25in]{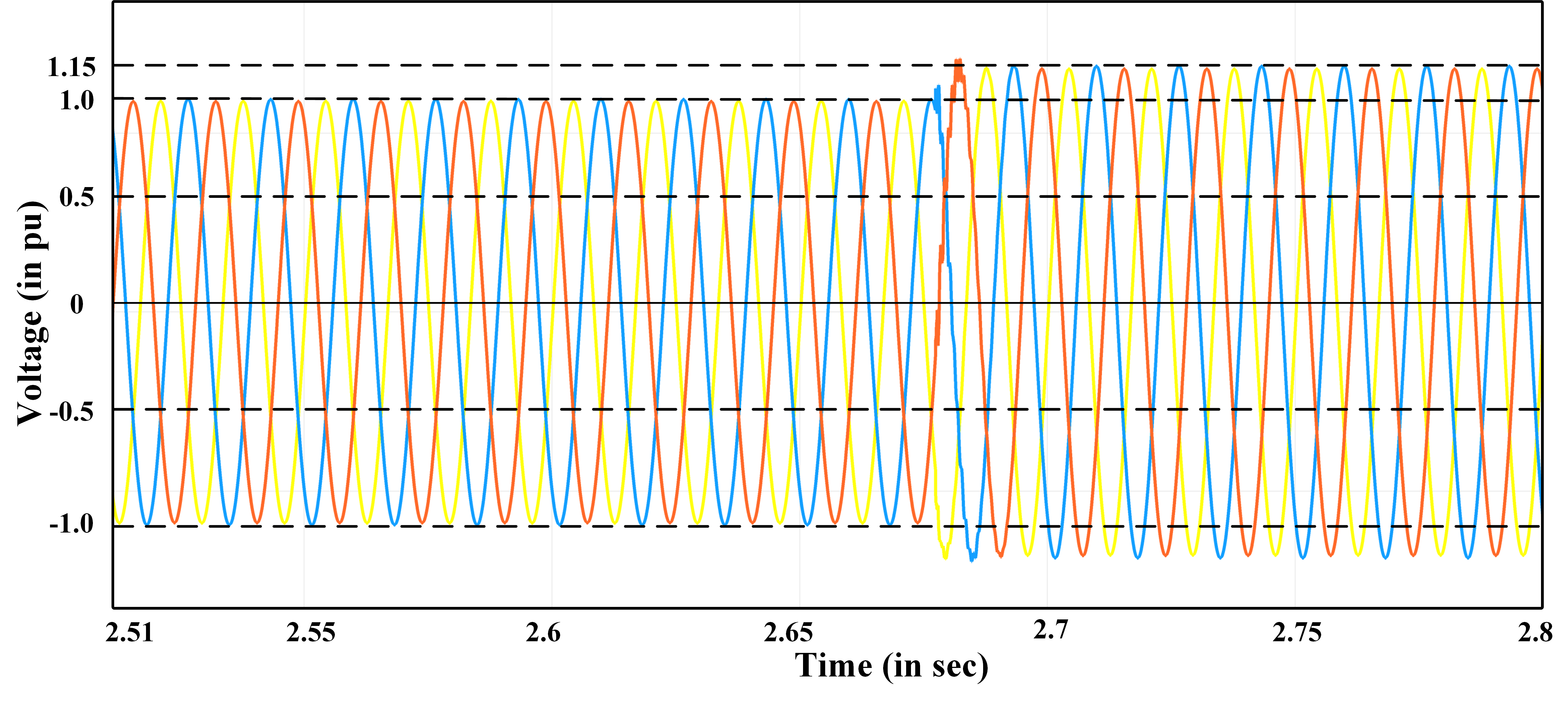}
\vspace{-1mm}
\caption{Voltage variation from $1.0$ p.u. to $1.15$ p.u. in distribution feeder at node 8 due to the packet modification attack \protect\circled{1}.}
\vspace{-3mm}
\label{unwanted_step}
\end{figure}

\vspace{-1mm}
\section{Results and Discussion}\label{s:results}
In order to investigate the impact of cyberattacks on the VVO scheme, the IEEE 34-bus radial system is considered in this paper. The 34-bus system has an operating voltage of $4.16$ kV. The system is modified to include VRs between nodes 7-8 and 19-20 as well as a PVSI on node 34, which is one of the optimal locations for DER placement in this benchmark \cite{9720991_DER_in_IEEE34}. The SI is implemented through the SI toolbox of \textsc{OPAL-RT} to enable Volt/VAR control. Moreover, the nodes of the feeder are monitored using smart meters which communicate voltages as well as active and reactive power to the master controller through DNP3 messages. The VVO is performed every 15 mins and changes in setpoints are updated. The modified network is modeled in \textsc{RT-LAB} and smart meters communicate through DNP3 slaves to \textsc{ExataCPS} through the \textsc{OpOutput} block of the \textsc{RT-LAB} library. The DNP3 master is a physically emulated IED running on a PC in the SITL setup, as shown in Fig. \ref{fig_1}. With the initial measurements, the VVO code computes the setpoints of VRs and the Volt/VAR curve of SI. These are written to the slave through direct operate messages by the master. Finally, the setpoints are received inside the simulation through the \textsc{OpInput} block of the \textsc{RT-LAB} library. A schematic representation of the modified 34-bus system with the \textsc{OpInput} and \textsc{OpOutput} blocks is shown in Fig. \ref{fig_3}. The results of the impact of cyberattacks on the VVO scheme in the developed setup are presented in two parts, one focusing on the analysis affecting the ADMS operation and the second focusing specifically in the effects on the CVR part.

\vspace{-2mm}
\subsection{Impact Analysis on ADMS Operation}


\begin{figure*}[!t]
\centering
\subfloat[]{\includegraphics[width=3.1in]{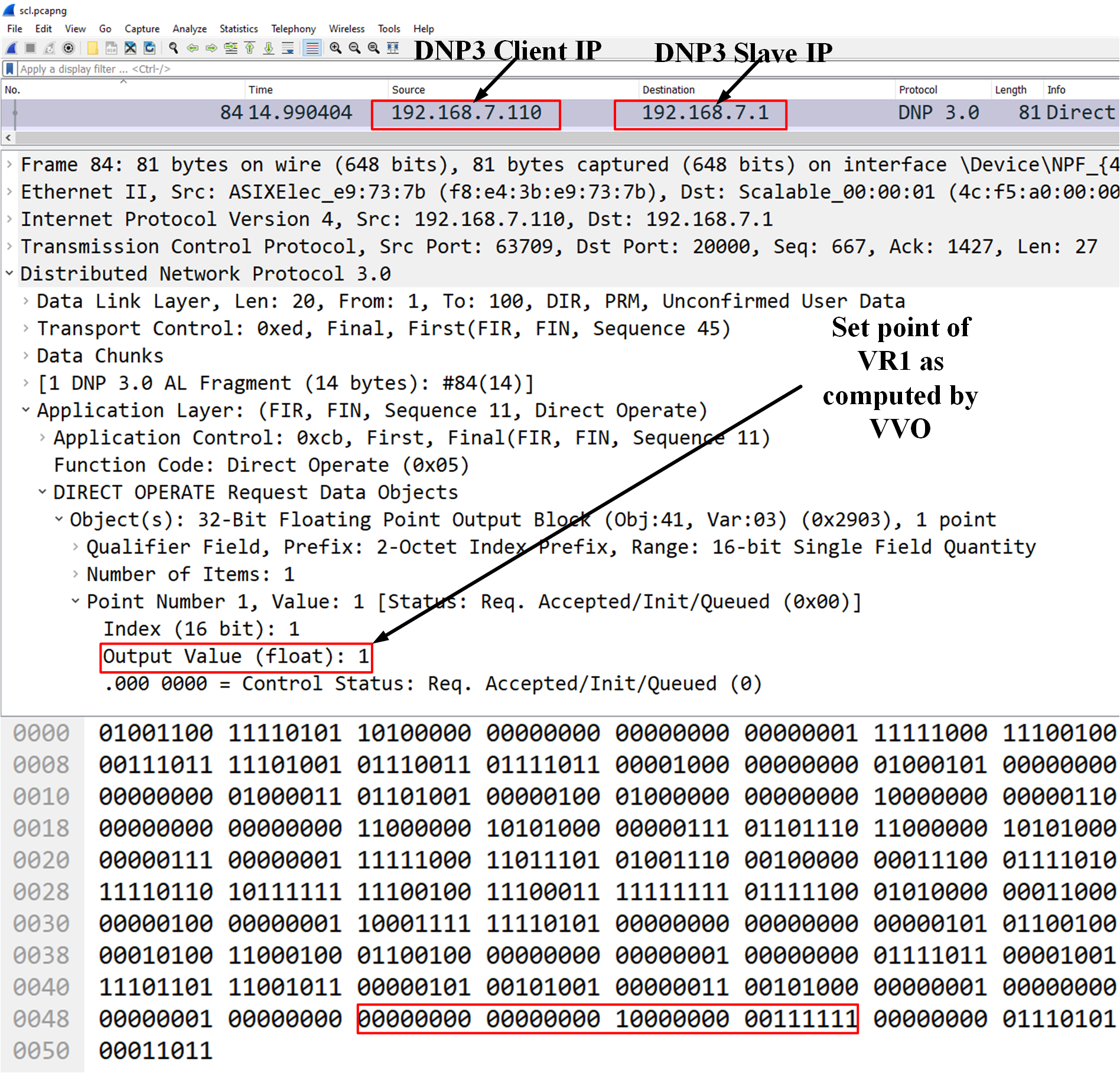}%
\label{fig_first_c}}
\hfil
\subfloat[]{\includegraphics[width=3.1in]{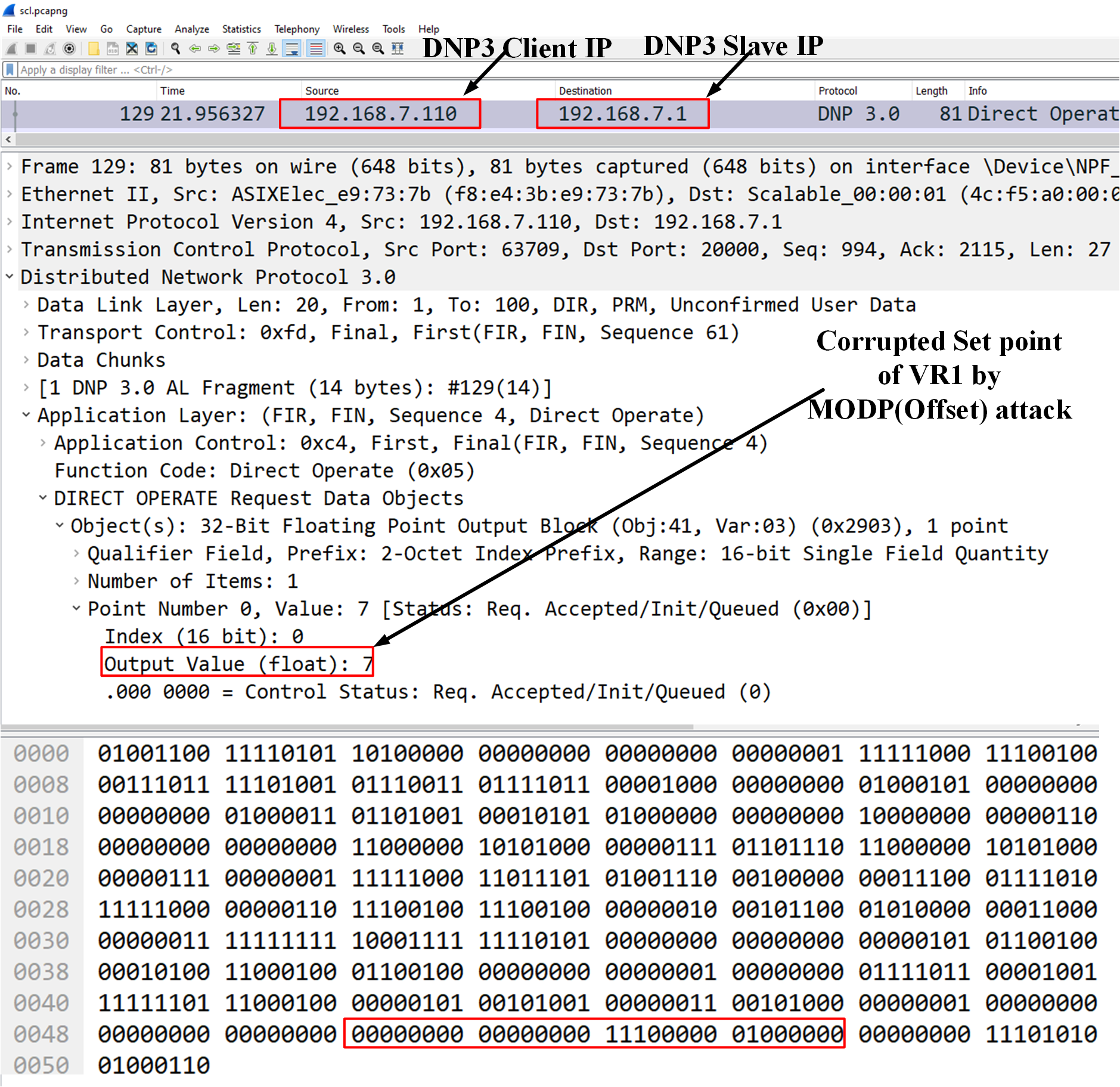}%
\label{fig_second_c}}
\caption{Wireshark capture of direct operate DNP3 message to set step value in VR1: (a) before, and (b) after, the add offset attack \protect\circled{1}.}
\vspace{-4mm}
\label{fig_5}
\end{figure*}

To illustrate the impact of attack category \circled{1}, we consider tampering the setpoint of VR1 during a load change operation. Let us assume a load change of 25\% on a balanced RL load with a capacity of $0.1$ MW and $0.6$ MVar at node 8 of the modified IEEE 34-bus system. In this scenario, the master controller sends a request to the DNP3 slave to retrieve values stored in the read buffer. The DNP3 slave acknowledges the master's request and sends a read response message, which includes measurement data of the load obtained from a smart meter. This information is then used to update the load change in the VVO algorithm and calculate optimized setpoints for VR1. Since the voltage remains within operational limits, the load increase does not require any voltage level adjustments. As a result, the setpoints of VR1 remain unchanged. Consequently, the DNP3 master sets the setpoint value of VR1 to $1$, reflecting the previous value of the VR1 step. However, in this scenario, the DNP3 client inside \textsc{ExataCPS} is subjected to a \emph{MODP} add offset attack, which modifies the payload of the DNP3 stream by adding an offset of $+6$. This alteration causes the communicated value to the DNP3 slave to increase by $6$ points, resulting in a corrupted setpoint for VR1 of $7$ instead of the intended 1. As a consequence, an undesired voltage increase is observed in the feeder near node 8, as illustrated in Fig. \ref{unwanted_step}. This voltage rise exceeds the operating voltage range defined by the ANSI standard, reaching $1.15$ p.u. To further analyze the impact of the attack, we examine the Wireshark captures of the DNP3 message both before and after the cyberattack, as shown in Fig.  \ref{fig_5}. It is evident that the MODP attack corrupts the setpoint of VR1, leading to an unintended change in voltage within the distribution feeder, as depicted in Fig. \ref{unwanted_step}.

\begin{figure}[t]
\centering
\includegraphics[width=3in]{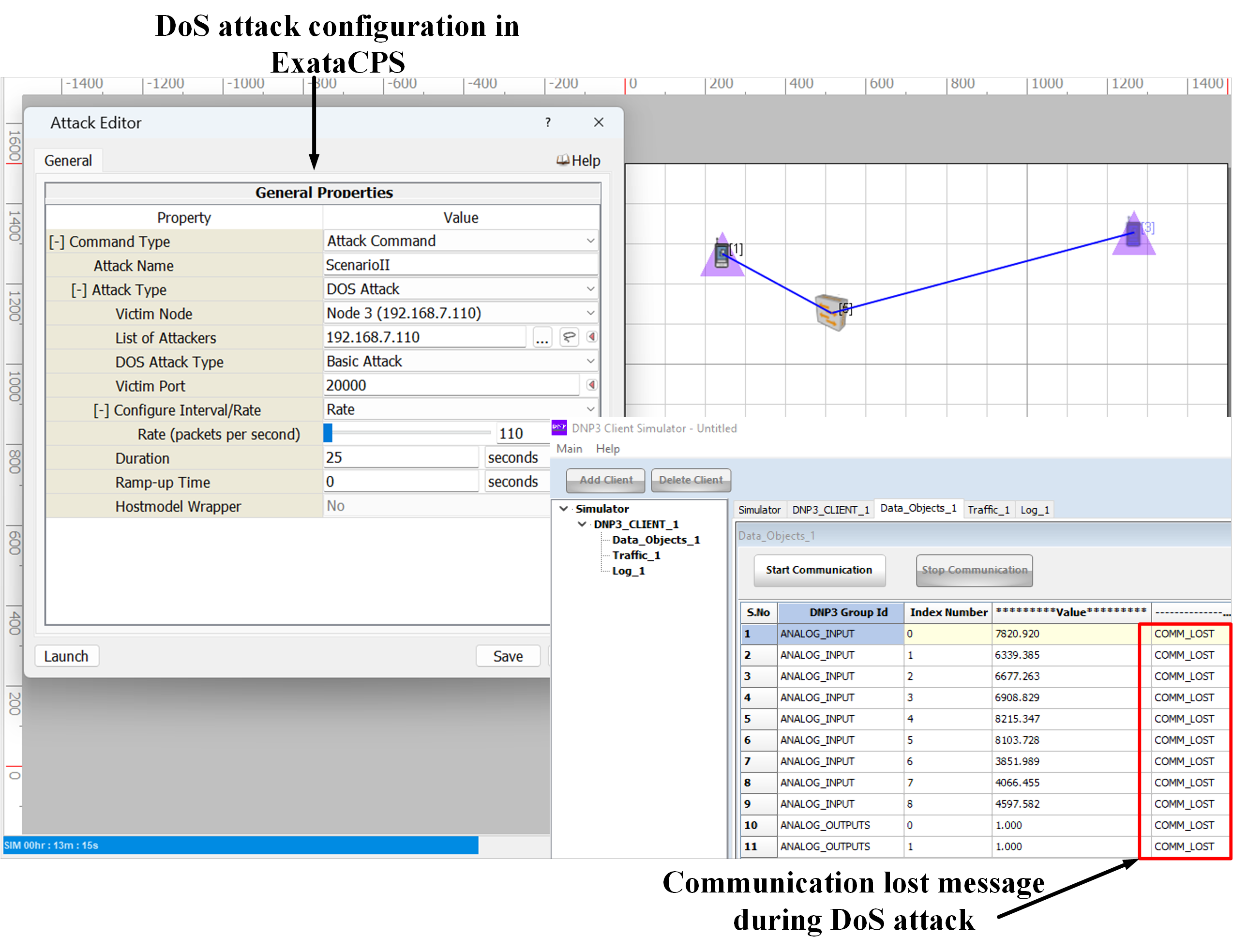}
\vspace{-1mm}
\caption{Once the DoS attack \protect\circled{2} is launched, the communication is lost to the DNP3 client.}
\vspace{-4mm}
\label{dos-scenario}
\end{figure}

The effect of attack category \circled{2}, i.e., cause the meter measurements not to be available for performing VVO, is demonstrated by launching a DoS attack on the DNP3 client in the \textsc{ExataCPS} interface. DoS attack works by flooding the victim node with excessive traffic thereby making it inoperative. Once the attack is launched, the DNP3 client simulator displays a connection lost status. As a result, the previously updated values of measurement messages are available with the master controller to run VVO for the next cycle, which in turn results in failure to compute updated setpoints for VVO, 
 thereby rendering VVO ineffective, as depicted in Fig. \ref{dos-scenario}. 

\begin{figure}[t]
\centering
\includegraphics[width=3.1in]{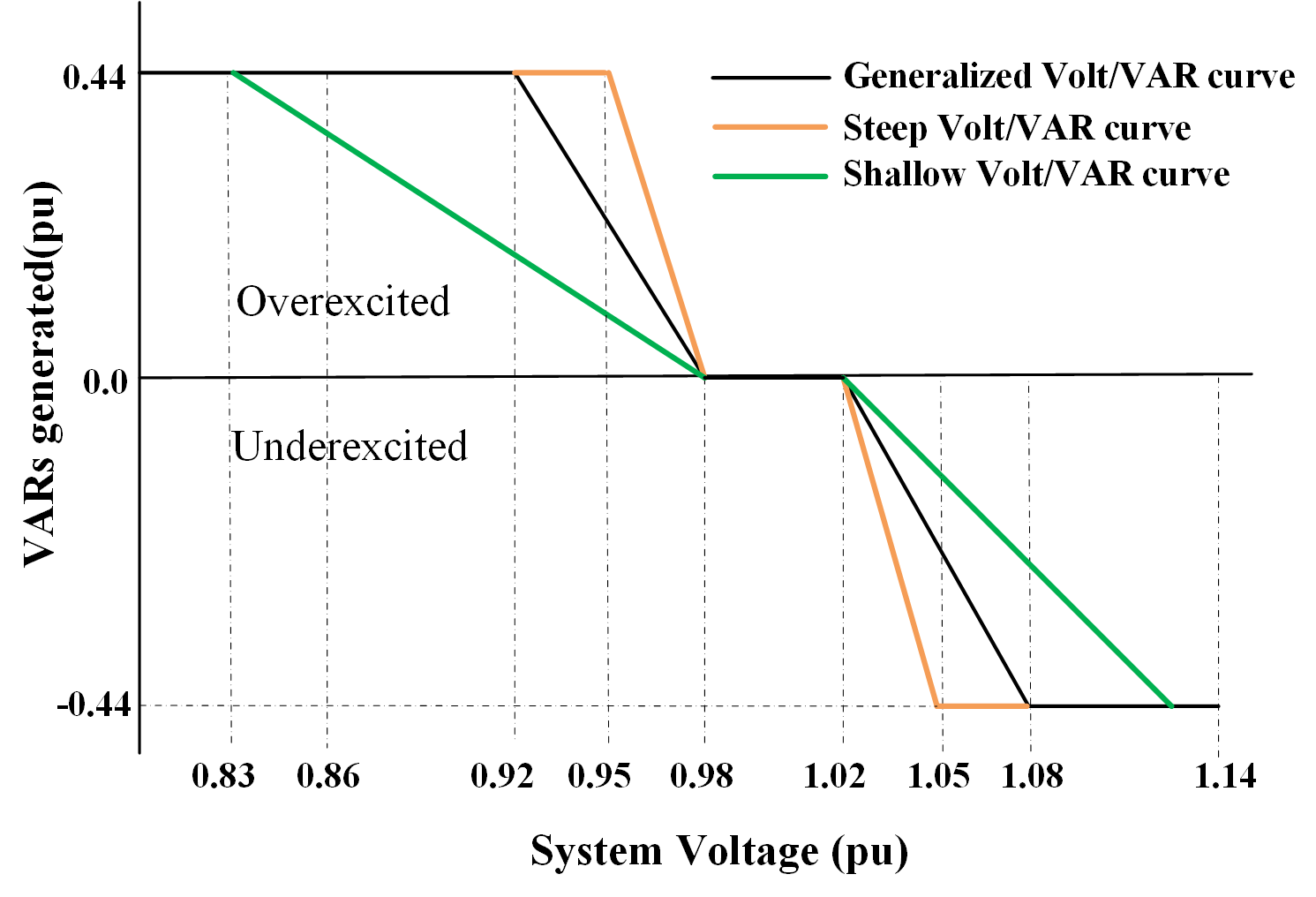}
\vspace{-1mm}
\caption{Modification of Volt/VAR curve during attack \protect\circled{3} on SI.}
\vspace{-4mm}
\label{fig_6}
\end{figure}

  \begin{figure}[t]
\centering
\includegraphics[width=3.1in]{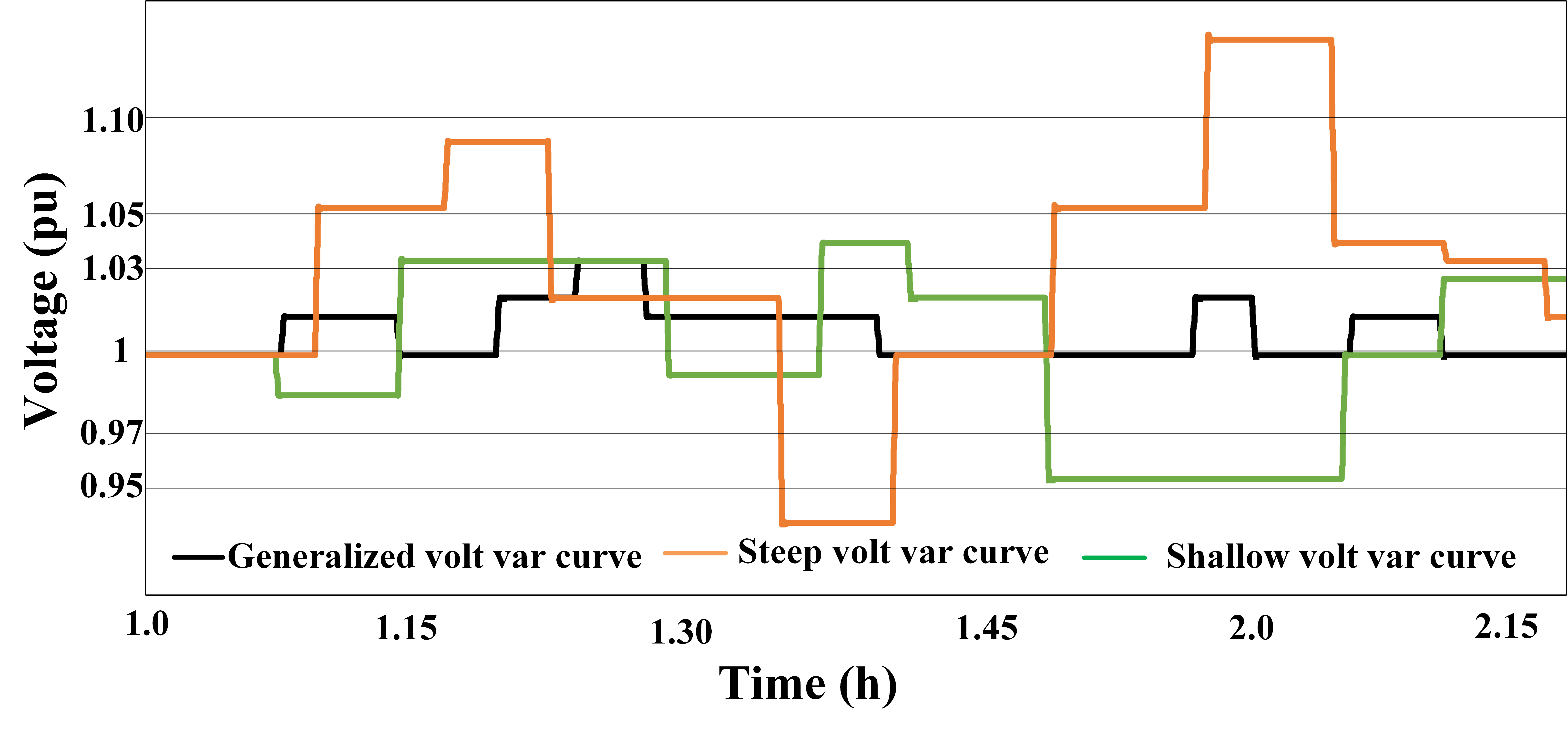}
\vspace{-2mm}
\caption{Voltage variations in voltage at node 34 due to modification in slope of Volt/VAR curve during attack \protect\circled{3} on SI.}
\vspace{-5mm}
\label{fig_7}
\end{figure}

The SIs provide reactive power support to the distribution network as per the generalized Volt/VAR curve regulated by IEEE 1547-2018 standard. In the case which the Volt/VAR curve setpoints of SI are modified under the attack \circled{3} scenario, the slope of the curve can be increased or decreased by an attacker resulting in a steep or shallow Volt/VAR curve, respectively, as shown in Fig. \ref{fig_6}. In \textsc{ExataCPS}, a MODP attack with a multiply option is selected to modify the slope of the Volt/VAR curve. The Volt/VAR curve setpoints are stored in an array and are updated in the secondary control block under the reactive power mask of the SI through \textsc{RT-LAB}. The node voltage at the PVSI node is monitored during the attack. Corrupted Volt/VAR curve settings lead to increased voltage transients as shown in Fig. \ref{fig_7}, and hence, this creates issues with the load, DER units, and switchgear components.

\subsection{Impact Analysis {on CVR}}
\vspace{-1mm}

In order to illustrate the impact of the attack scenarios on CVR, we run the simulation for a 24-hour period in which the MODP attack realization is launched using \textsc{ExataCPS} library. During the simulation, the changes in the number of setpoints in VRs and Volt/VAR curve are measured. Using this information, the energy savings achieved through CVR are calculated. Subsequently, the results are compared with the energy savings achieved in the absence of an attack.

The load demand profile for a day is considered for the simulation as in \cite{ShailendraEventdrivenPV_9350202}. It is observed that the load consumption is lower during the night and early morning compared to the daytime. Thus, the load profile is divided into two stages: a lightly loaded and a heavily loaded period. The average electricity purchase rate, denoted as ${C_{E,t_k}}$, is considered as \$$0.20$ per kWh for a day \cite{Electric_Price}. The operating cost of VR is considered as \$$0.05$ per tap change \cite{VVO_operating_cost_MANBACHI201628}. The CVR factor, which represents the ratio of percentage energy saved to percentage reduction in voltage achieved through CVR, is computed for each attack following the method in \cite{CVR}. A higher value of CVR factor indicates higher savings and vice-versa.

Table \ref{Table I} presents the total number of updates in setpoints for the VRs under the different attack categories. Under \emph{tampering VVO setpoint} attack \circled{1}, increased number of setpoint updates in VRs occurs as compared to normal operation. When a DoS attack is launched under \emph{tampering smart meter measurements} attack \circled{2}, communication between VVO controller and VRs is lost and hence no changes in setpoint of VRs are observed. Therefore, results for this attack are not included in Table \ref{Table I}. In the case of  \emph{modifying SI Volt/VAR curve} attack \circled{3}, when the slope of the Volt/VAR curve is reduced resulting in a shallow curve, increased number of setpoint updates for VRs occurs. A reduced number of setpoint updates for VRs is observed under steep Volt/VAR curve, however, this leads to an increased number of voltage limit violations as shown in Fig. \ref{fig_7}.

\begin{table*}[htbp]
  \centering
  \caption{Number of setpoint updates in VRs during the simulation.}
   \resizebox{\textwidth}{!}{
  \begin{tabular}{ccccccccc}
    \toprule
    Operating Devices & \multicolumn{2}{c}{Normal operation} & \multicolumn{2}{c}{\protect\circled{1} Tampering VVO setpoints} & \multicolumn{4}{c}{\protect\circled{3} Modifying SI Volt/VAR curve setpoints} \\
    \cmidrule(r){2-3} \cmidrule(lr){4-5} \cmidrule(lr){6-9}
    & Light load & Heavy load & Light load & Heavy load & \multicolumn{2}{c}{Shallow Curve} & \multicolumn{2}{c}{Steep Curve} \\
    \cmidrule(r){6-7} \cmidrule(lr){8-9}
    & & & & & Light load & Heavy load & Light load & Heavy load \\
    \midrule
    {VR1} & $4$ & $7$ & $16$ & $24$ & $5$ & $10$ & $3$ & $5$ \\
    VR2 & $9$ & $12$ & $21$ & $17$ & $11$ & $16$ & $8$ & $9$ \\
    \bottomrule
  \end{tabular}
  }
  \label{Table I}
\end{table*}

Table \ref{Table II} presents the cost analysis and CVR factor for the attack categories. 
The procedure for calculating the cost terms in Table \ref{Table II} is as follows. The average electrical energy supplied from the substation in the modified IEEE 34-bus system is measured in \textsc{RT-LAB}. This is then multiplied by the average electricity purchase rate to obtain the energy purchase cost. The operating cost of the VRs is calculated by multiplying the number of setpoints updates in Table \ref{Table I} with the cost per tap change. The total active power loss is computed using the modified ADMM algorithm, and then used to calculate energy loss cost. The total operational cost, which includes energy cost, losses, and the operating cost of the VRs, is compared to a baseline case in which CVR is not implemented and no attack occurs. This serves as the reference point for evaluating the percentage cost savings. 
The first two columns of Table \ref{Table II} correspond to values when VVO is implemented without and with CVR, respectively. The other columns correspond to cost values for attack categories \circled{1} and \circled{3}, respectively. As for attack \circled{2}, due to a successful DoS attack, communication was lost, and thus, no outcomes could be obtained. 
Based on the presented results in Table \ref{Table II}, it can be concluded that implementing CVR with VVO and without any attack being realized yields cost savings of $3.17$\% and $1.44$\% during light load and heavy load conditions, respectively. The cost savings as well as the CVR factor deteriorate under the effect of attacks. It is worthwhile to mention that although under Volt/VAR curve with steep slope attack, cost savings are better as compared to other attack categories, this scenario leads to a higher number of violations in voltage limits as per the ANSI standard C84.1-2011 (Fig. \ref{fig_7}).

\begin{table*}[htbp]
  \centering
  \caption{Cost analysis and CVR under the different attack scenarios.}
   \resizebox{\textwidth}{!}{\begin{tabular}{ccccccccccc}
    \toprule Cost terms
    &
    \multicolumn{2}{c}{VVO without CVR} & \multicolumn{2}{c}{VVO with CVR} & \multicolumn{2}{c}{\protect\circled{1} Tampering VVO setpoints} & \multicolumn{4}{c}{\protect\circled{3} Modifying SI Volt/VAR Curve Setpoints} \\
    \cmidrule(r){2-3} \cmidrule(r){4-5} \cmidrule(r){6-7} \cmidrule{8-11}
    & Light load & Heavy load & Light load & Heavy load & Light load & Heavy load & \multicolumn{2}{c}{Shallow Curve} & \multicolumn{2}{c}{Steep Curve} \\
    \cmidrule(r){8-9} \cmidrule(r){10-11}
    & & & & &  & & Light load & Heavy load & Light load & Heavy load \\
    \midrule
    Energy purchased (\$) & $7288$ & $12776$ & $6873$ & $12478$ & $7117$ & $12506$ & $7137$ & $12601$ & $7208$ & $12253$ \\
    VR operating costs (\$) & -- & -- & $0.65$ & $0.95$ & $1.85$ & $2.05$ & $0.8$ & $1.3$ & $0.55$ & $0.7$\\
    Cost of Energy loss (\$) & $97$ & $161$ & $63$ & $113$ & $91$ & $179$ & $79$ & $153$ & $76$ & $152$\\
    \% Energy saved & -- & -- & $3.17$ & $1.44$ & $1.07$ & $0.6$ & $0.9$ & $0.1$ & $1.41$ & $2.8$\\
    CVR factor & -- & -- & $0.8$ & $0.57$ & $0.36$ & $0.3$ & $0.23$ & $0.46$ & $0.23$ & $0.46$\\
    \bottomrule
  \end{tabular}}
  \label{Table II}
  \vspace{-4mm}
\end{table*}

\section{Conclusion}\label{s:conclusions}
This paper investigates the impact of cyberattacks on the VVO scheme using a real-time co-simulation setup. A detailed discussion of the setup considered for this study is provided. 
The developed co-simulation setup is flexible and can be seamlessly adapted to other industrial communication protocols used in electric utilities such as IEC 61850, OpenADR, IEEE 2030.5 etc. 
The results demonstrate that the number of setpoint updates of VRs increases under influence of cyberattacks. Moreover, cyberattacks on modifications of VVO device setpoints lead to an increased number of voltage limit violations on the distribution feeder. CVR factor also reduces under the influence of the attacks. The benefit of the developed real-time co-simulation setup is that it can be employed to assess the impacts of different attack scenarios and to test mitigation strategies before field deployment.

\vspace{-2mm}
\bibliographystyle{IEEEtran}
\bibliography{bibi.bib}

\newpage

\end{document}